# Ozone mini-hole in winter 2015/2016 over Siberia: Characteristic features and relationship to atmospheric blocking

Mokhov I.I.[1,2], Sitnov S.A.[1]

[1]A.M. Obukhov Insitute of Atmospheric Physics RAS

[2]Lomonosov Moscow State University

mokhov@ifaran.ru

**Abstract**

The characteristic features and mechanisms of the formation of a deep ozone mini-hole (OMH) in winter 2015/2016 over Siberia were carried out using AIRS and CALIOP satellite data. The depletion in the total column ozone was caused mainly by a strong decrease in ozone mixing ratio in the lower stratosphere, reaching 50% near the pressure level of 70 hPa. The formation of OMH was closely related to the dynamic factors including an increase in the tropopause height and advection in the troposphere of ozone-depleted subtropical air, which were associated with atmospheric blocking. Also, an important factor of OMH formation was ozone decrease in the stratosphere due to increased outflow of ozone from the area of the OMH. The contribution of chemical destruction of stratospheric ozone, due to catalytic reactions with halogen-containing compounds on the surface of polar stratospheric clouds to the OMH formation, was evaluated.

**Keywords:** ozone, mini-hole, atmospheric blocking, polar stratospheric clouds, winter 2015/2016, Siberia.

## 1. Introduction

A deep synoptic-scale anomaly, manifesting in total column ozone (TCO) field, is considered as an ozone mini-hole (OMH) (Newman et al., 1988; James et al., 1997; James, 1998). OMHs are revealed in middle and high latitudes, more often in winter than in summer and more often in the Northern Hemisphere than in the Southern Hemisphere (James, 1998). In contrast to the Antarctic ozone hole, the formation of OMH is mainly due to dynamical factors - horizontal northward advection of ozone-depleted subtropical air and upward movements of air in the upper troposphere and lower stratosphere (Peters et al., 1995; Petzod, 1999; Iwao and Hirooka, 2002), however, the possible contribution of chemical ozone destruction, conditioned by heterogeneous chemical reactions on the surface of polar stratospheric clouds (PSCs) is also discussed (Petzod, 1999; Teitelbaum et al., 2001; Timofeev et al., 2020). The importance of studying the chemical mechanism of the OMH formation is justified due to a systematic increase in PSC occurrence during Arctic winter (Tritcher et al. 2021).

There are no generally accepted criteria for the OMH determining. As thresholds, both absolute TCO values (usually 220 Dobson units (DU), which is the threshold accepted for determination of the Antarctic hole) and also the TCO anomalies with values in the range from -50 to -70 DU, calculated with respect to different climatological averages, are taken (Millan and Manney, 2017).

The formation of OMH is often accompanied by blocking of the zonal flow in the mid-latitude troposphere. According to Barriopedro et al. (2010), half of the OMHs over Europe is manifested under atmospheric blocking conditions, moreover this kind of OMHs is characterized by the greatest depth and enhanced stability. Teitelbaum et al. (2001) concluded that the synoptic-scale dynamics, which is the primary mechanism for the OMH formation, contribute greatly to the PSC formation the Arctic. Due to continued global warming, against the background of an increase in the likelihood of the prolonged atmospheric blocking events (Mokhov and Timazhev, 2019), an increase in the frequency of winter OMHs is expected in the Euro-Atlantic sector.

The physical mechanisms responsible for the formation of OMH during anomalously prolonged summer 2010 atmospheric blocking event over European Russia (ER) were analyzed by Sitnov and Mokhov (2015, 2016). It was shown that the decrease in TCO within the block was mainly due to the peculiarities of regional atmospheric circulation associated with omega-block. TCO modulates the intensity of solar UV radiation, reaching the Earth's surface (Madronich et al., 1998). In the summer time because of the high Sun position and cloudless anticyclonic weather, a decrease in TCO can result the living things within blocking regions will be exposed by hazardous dose of biologically active solar UV radiation. Thus, ozone depression in the north of ER in summer 2010 was accompanied by 23% increase in the daily dose of erythemal solar irradiation (Sitnov and Mokhov, 2019). In winter, due to the low position of the Sun over the horizon, an increase in solar UV radiation within the blocking area poses less danger. On the other hand, winter OMHs can be much deeper than summer one (James 1998).

One of the deepest OMH in the atmosphere of Russian Federation was detected over Western Siberia in winter 2015/2016. Despite this OMH was the subject in a number of studies, its mechanisms are still not entirely clear. Nikiforova et al. (2018) explained the OMH formation by the dynamical processes inside the polar vortex. According to Timofeev et al. (2018), the short-term TCO variability associated with the OMH dynamics was hardly induced by photochemical ozone destruction resulted from heterogeneous halogen activation on PSCs. Sitnov and Mokhov (2021) pointed on the relationship of the OMH with peculiarities of regional atmospheric circulation, conditioned by atmospheric omega-block. The purpose of this work is a comprehensive analysis of the characteristics and mechanisms of the OMH over Western Siberia in winter 2015/2016, including an analysis of the relationship of the OMH with a concomitant atmospheric blocking event.

**2. Data used**

To study the characteristics of the OMH, ozone measurements from the AIRS satellite instrument, aboard the Aqua platform were used. AIRS (Atmospheric InfraRed Sounder) retrieves the wide range of the profiles of various meteorological parameters and atmospheric gases using nadir geometry of measurements (Auman et al., 2003). The AIRS ozone retrieval algorithm uses measurements of radiance in the wavelength range of 9.3–10.1 μm and (up to September 2016) the accompanying measurements of the AMSU (Advanced Microwave Sounding Unit) microwave radiometer, that make it possible to retrieve atmospheric parameters at cloud coverage up to 80% (Susskind et al., 2006). The aggregate spatial resolution of AIRS/AMSU retrieval in nadir is ~45 km. The calculated accuracy of determining the total ozone content in the atmosphere is 20%. Air temperature and geopotential heights at different pressure levels as well as tropopause characteristics, obtained by the AIRS instrument, have been also involved in the analysis. The temperature measurement accuracy is calculated to be 1K in a 1 km layer (Auman et al., 2003). Daily daytime measurements of AIRS, available at 24 levels in the atmosphere in the range of 1000–1 hPa on a latitude–longitude grid 1° × 1° (AIRXSTD 007) as well as calculated total column ozone, were obtained using visualization and analysis system Giovanni (https://disc.sci.gsfc.nasa.gov/giovanni) developed and maintained by NASA GES DISC (Acker and Leptoukh, 2007).

The peculiarities of regional atmospheric circulation under atmospheric blocking conditions were studied using data from NCEP/NCAR reanalysis (Kistler et al., 2001). The data are available on 17 mandatory levels (from 1000 hPa to 10 hPa) provided on 2.5° × 2.5° latitude-longitude grids. The daily horizontal wind components as well as geopotential heights at different pressure levels were obtained with the help of http://www.esrl.noaa.gov/psd/data/reanalysis server. The long-range air mass transfer in the OMH region in the troposphere and stratosphere was studied using the matrix type of trajectories, calculated with the help of the NOAA HYSPLIT Lagrangian atmospheric transport and dispersion modeling system (Stein et al, 2015), using meteorological data from GDAS with 1°×1° spatial resolution. The backward trajectories' ensembles were obtained with the help of https://www.ready.noaa.gov/HYSPLIT_traj.php server.

To assess the contribution of the chemical destruction of ozone within the ozone mini-hole, the CALIOP L2 PSC mask products were used. CALIOP (Cloud-Aerosol Lidar with Orthogonal Polarization) is a near-nadir viewing two-wavelength (532 and 1064 nm) lidar, installed onboard the CALIPSO satellite (Hunt et al., 2009). CALIOP retrieves vertical profiles of aerosols and clouds along 14–15 orbits per day in the latitudinal belt from 82° S to 82° N with the high spatial resolution (5-km along-track × 180-m vertical) over altitudes from 8.4-30 km (Winker et al., 2009). The 532-nm signal is separated into orthogonally polarized components ($\beta_\perp$ and $\beta_\parallel$). Using $\beta_\perp$ and scattering ratio $R_{532}=(\beta_\perp+\beta_\parallel)/\beta_m$ ($\beta_m$ - molecular backscatter at 532 nm) the CALIOP algorithm allows identifying seven tropospheric aerosol subtypes and four stratospheric ones, including PSC aerosol. Moreover, using the measured aerosol depolarization ratio ($\beta_\perp/\beta_\parallel$) and inverse scattering ratio $1/R_{532}$, CALIOP v2 PSC algorithm (Pitts et al., 2018) allows classifying PSCs by composition (Pitts et al., 2013; Tritscher et al., 2021). Because of low signal-to-noise ratio the PSC studies are based mainly on night-time measurements. The CALIPSO Lidar PSC data (CAL_LID_L2_PSCMAsk-Prov-V1-10_V1-00) were taken through the online NASA server at the address https://www-calipso.larc.nasa.gov.

Daily AIRS daytime measurements available at 24 levels in the atmosphere in the range of 1000–1 hPa on a latitude–longitude grid 1° × 1° (AIRXSTD 007), were obtained using visualization and analysis system Giovanni (https://disc.sci.gsfc.nasa.gov/giovanni) developed and maintained by NASA GES DISC (Acker and Leptoukh, 2007).

**3. Methods**

The study widely uses calculations of the anomalies of atmospheric parameters averaged over the area or/and in time, manifesting under atmospheric blocking conditions during winter 2015/2016. The 5-day period from January 26 to January 30, 2016 was characterized by strengthening of the blocking anticyclone and thus was chosen as a reference blocking period. It is worth noting that the estimates of the anomalies based on the average values are more statistically robust.

Spatial distributions of anomalies of atmospheric parameter X ($\Delta X$) were calculated as deviations of the local mean values of X (in the resolution 1°×1°) for January 26-30, 2016 from the corresponding long-term averages of X, calculated during the periods January 26-30 of 2003-2015.

Latitude -pressure distributions of $\Delta X$ were calculated as deviations of the local mean values of X at each latitude (step 1°) and at each of the 24 pressure levels for January 26-30, 2016 from the corresponding long-term means of X during the periods January 26-30 of 2003-2015.

Time-pressure distributions of ΔX were calculated for each day and at each of the 24 pressure levels as the deviations of X values averaged over the atmospheric blocked region (60°-75°N, 70°-100°E) in 2916, from the corresponding to a given day and the pressure level the long-term means of X in the 2003-2015, averaged over the same territory.

Daily regional anomalies of X were calculated as the deviation of daily spatial-averaged values of X in 2016 from corresponding to a given day the long-term means of X during the period from 2003 to 2015, averaged over the same territory.

**4. Results and discussion**
*4.1. Large-scale circulation in the atmosphere of the Northern Hemisphere in winter 2015/2016*

The extratropical large-scale atmospheric circulation in the Northern Hemisphere in the winter of 2015/2016 possessed a number of distinctive characteristics. In the troposphere, in early January 2016, an intensification of wave formation processes was started that led to the establishment of a regime of stationary planetary Rossby-type waves, characterized by a wavenumber k = 3 with a subsequent decrease in the wavenumber to k = 2 at the end January. During most of winter 2015/2016, the wave crest of the highest amplitude, was positioned over the north of Western Siberia.

A deep anomaly-cold stratospheric polar vortex (SPV) centered near the North Pole was formed in early September 2015 and existed until mid-January 2016. At the end of January 2016, the SPV was stretched out along the meridians 120°E - 300°W with the formation of two low-pressure centers, with one located over Greenland, and the deeper one located over the north of the Central Siberian Plateau. Such a configuration of SPV favored to the development of a planetary wave with the wavenumber k = 2 in the stratosphere. It is characteristic that at the end of January 2016, planetary waves in both the troposphere and the stratosphere were characterized by the same wave numbers (k = 2).

*4.2. Relationship between total ozone fluctuations and atmospheric blocking over Northern Eurasia in the winter of 2015/2016.*

The wave crest anchored over Western Siberia coincided with the periphery of a closed high-pressure area (an anticyclone), that embraced the entire troposphere. The manifestation of strong and extensive stationary anticyclone, which impedes (blocks) the western transport of air masses in the mid-latitude troposphere, is called atmospheric blocking (Rex 1950). Fig. 1a shows the results of daily diagnostics of atmospheric blocking in the longitudinal segment 55°-140°E for winter 2015/2016. For the diagnosis of the blocking episodes, the 500-hPa geopotential height ($H_{500}$) data were analyzed, using the blocking index similar to that proposed by (Tibaldi and Molteni, 1990). For each day and each longitude, the northern ($G_N$) and southern ($G_S$) meridional gradients of $H$500 were calculated using:

$$G_N = \frac{H500(\varphi_N) - H500(\varphi_0)}{\varphi_N - \varphi_0},$$

$$G_S = \frac{H500(\varphi_0) - H500(\varphi_S)}{\varphi_0 - \varphi_S},$$

where $\varphi_N = 80°$ N $+ \delta$, $\varphi_0 = 60°$ N $+ \delta$, $\varphi_S = 40°$ N $+ \delta$, $\delta = -5°, 0°, 5°$. In a separate day a given longitude is defined as "blocked" if the following conditions are satisfied for at least one value of $\delta$:

$$G_S > 0,$$

$$G_N < -10 \text{ m per latitude degree}.$$

The definition of atmospheric blocking often additionally includes the conditions that blocking persists for at least five days and a NH blocking anticyclone during its lifetime situates poleward from 35° N and also that the ridge exceeds 5° latitude (Lupo and Smith, 1995). It can be seen from Fig 1a that there were three basic atmospheric blocking episodes, occurred over considered North Eurasian region in winter 2015/2016 - the first one in early January (1.01.2016-6.01.2016), the second one at the end of January (19.01.2016-30.01.2016) and the third one in mid-February (11.02.2016-21.02.2016).

Fig 1b shows the winter 2015/2016 day-to-day variations in TCO averaged over the region of North Eurasia, restricted by the coordinates 55°–85° N and 55°-140 °E. It can be seen from a comparison of Fig.1a and Fig. 1b that each blocking episode is followed by a decrease in the regional total column ozone.

*4.3. Spatial and vertical distributions of ozone anomalies over the blocked region*

According to the NCEP/NCAR reanalysis data the spatial distribution of $H$500 over Siberia in the second half of January 2016 (Fig 2a) indicates manifestation an extensive quasi-stationary high-pressure system with the maxima of $H$500 reaching 5500 gpm (geopotential meters) near 60° N, 80° E and two adjacent low-pressure regions located to the west and to the east from it that is rather typical for an atmospheric block of omega type. Joint analysis of $H$500 and TCO fields indicates the formation of a deep regional anomaly of ozone coinciding with the blocked region (cf. Fig. 2a and Fig. 2b). The local (1° × 1°) five-day average TCO anomalies over the north of Western Siberia in the end of January 2016 and in the beginning of February reached -140 DU (-40% of corresponding long-term mean values), while the area with negative TCO anomalies less than -100 DU exceeded 1 million km$^2$.

At the end of January 2016, the latitude-pressure structure of ozone VMR anomalies in the longitude sector of OMH manifestation, indicates a decrease in the ozone content over middle and high latitudes of the Northern Hemisphere (Fig.3a). North of 45° N, ozone decreases in the troposphere (stratosphere) averaged -4(-26)%, respectively. Extremal negative ozone anomalies manifest themselves in the 62°-69° N latitudinal belt in the vicinity the pressure level of 70 hPa.

*4.4. Physical mechanisms of ozone mini-hole over Siberia in winter 2015/2016*

In a free atmosphere, the large-scale air motions tend to follow isohypses. Analysis of the wind field at the 500-hPa pressure level indicates, that the air flow, when approaching the omega block, deviated to the north, skirted the blocking area, and returned again to the south (Fig. 4a). Due to increase of the ozone concentration from south to north in the upper troposphere and lower stratosphere, the southern flow at the western periphery of the blocking anticyclone leads to the replacement of ozone-enriched air by ozone-depleted air, that results in the decrease of TCO within the blocked region. The replacement of the mid- and high-latitude air by the subtropical air is also associated with a rise in the tropopause height over the blocked region (Fig. 4b), that also decreases TCO in this region. The analysis results show that in late January 2016, spatial anticorrelation between $\Delta O_3$ and $\Delta H_{trop}$ amounted to $r = -0.61$ over the region (0°-160° E, 25°-85° N) and reached $r = -0.70$ over the region (0°-160° E, 45°-85° N) (cf. Fig. 2a with Fig. 4b).

A negative spatial correlation between $\Delta O_3$ and $\Delta H_{trop}$ can be explained by vertical air motions embracing the upper troposphere and lower stratosphere. A fairly rapid rise of air is associated with its adiabatic cooling and also with an increase of the tropopause height. The accompanying rising of air in the lower stratosphere, leads to a decrease (increase) in the $O_3$ concentration below (above) the ozone maximum, respectively. Due to a decrease in the ozone photochemical relaxation time with altitude, the concentration of ozone in the upper part of the ozone profile, coming in a state of photochemical equilibrium, decreases faster than it increases in the lower part of the profile, which causes a decrease in TCO. On the contrary, a sufficiently rapid air descent leads to a decrease in the tropopause height and an increase in TOC. It should be noted that ascending motions in the atmosphere over the western and northern peripheries of the blocking anticyclone over European Russia in the summer of 2010 were diagnosed earlier in Sitnov et al. (2017).

Atmospheric blocking is associated with the establishing and amplification of stationary planetary waves (Naujokat and Pawson, 1996). Matsuno (1980) shown that with the wave crests (deviations of the mean zonal air flow to the north) the ascending air motions are associated, while troughs are associated with descending air motions. As it was noticed above, the upward air motions contribute to a decrease in TCO. Therefore, the formation of a negative TCO anomaly in the blocked region could facilitated the ascending air motions in the wave crest over the northern periphery of the blocking anticyclone (cf. Fig. 2a with Fig. 2b).

Figs. 5 demonstrate the relationship between TCO within OMH with the peculiarities of regional atmospheric dynamics associated with the atmospheric blocking event. The day-to-day variations in the TCO averaged over the blocked region (60°-75° N, 70°-100° E) are shown in Fig 5a, while Fig. 5b represents the corresponding variations of meridional wind in the free atmosphere, averaged along the western periphery (50°-75° N, 55°-70° E) of the blocking anticyclone ($v_{S \to N}$), and the eastern periphery (50°-75° N, 95°-110° E) of the block ($v_{N \to S}$). The results of the correlation analysis of the synoptic variations (calculated using detrended data) indicate that the variations of $v_{S \to N}$ and $v_{N \to S}$ under the blocking conditions were characterized by statistically significant anticorrelation: $r = -0.68$; 95% CI: (-0.84, -0.41). The connectedness of meridional wind flows indicates that the atmospheric circulation inside the anticyclonic vortex was nearly closed. It can be seen from comparing Fig. 5a with Fig. 5b that during the period of atmospheric blocking, the intensification of anticyclonic circulation was accompanied by a decrease in the TCO in the blocked region. It is characteristic, that the negative correlation of variations in TCO over the blocking area with the $v_{S \to N}$ amounts to $r = -0.67$ (-0.84, -0.39) and appears to be statistically significant.

Fig. 5a also shows the day-to-day variations in the tropopause height ($H_{trop}$) over the blocked region. It can be seen from comparing Fig. 5a with Fig. 5b that under blocking conditions, strengthening of the meridional wind over the western periphery of the block, was accompanied by an increase in the tropopause height over the blocked region. This fact indicates that day-to-day variations in the tropopause height over the block were associated with atmospheric dynamics. The correlation analysis of the synoptic $O_3$ and $H_{trop}$ variations indicates a statistically significant relationship between TCO and $H_{trop}$: $r = -0.65$ (-0.83, -0.36). Based on a linear regression model, the estimate of sensitivity of the TCO changes to the changes in $H_{trop}$ during the period from January 15 to February 10, 2016, was obtained to be -28 DU / km. This result is in tune with the results of Hoinka et al. (1996), concluded that 50% of the total columnar ozone variations can be explained by variations in the tropopause pressure. Since about half of the variance (48%) of the tropopause height daily variations could be associated with the meridional advection, it is difficult to separate the contributions of quasi-horizontal advection and variations of the tropopause in the formation of the OMH. Nevertheless, the presented quantitative estimates indicate a significant contribution of the tropospheric dynamic processes to the formation of the negative ozone anomaly over Siberia in 2015/2016 winter.

*4.5. Trajectory analysis*

The results of the trajectory analysis indicate that the sources of air masses, arriving into the blocked region in the troposphere and stratosphere were fundamentally different (Fig. 6). In the stratosphere, the trajectories of air parcels arriving in the region over the block were originated within the polar vortex, which was located at that time near the North Pole. Because the long-term latitudinal distribution of ozone in the stratosphere above the longitudinal segment 60°-110° E decreases from south to north (Fig. 3b), the transfer of ozone-depleted air from the polar stratosphere to the south contributed to a decrease in TCO over the region of atmospheric blocking. On the contrary, in the troposphere, air masses arrived in the blocking area from the south direction. Since the long-term latitudinal distribution of ozone in the troposphere in the abovementioned longitudinal segment is characterized by an increase from south to north (Fig. 3b), the transport of ozone poor air from the south also contributed to a decrease in TOC over the blocked region.

Using ozone volume mixing ratio (VMR) profiles obtained by the AIRS instrument and the wind data from the NCEP / NCAR reanalysis quantitative estimates of the ozone inflow into the blocked region at different altitudes were obtained The meridional ozone fluxes were calculated on the basis of the average values of the ozone VMR ($r_3$), air temperature and meridional wind profiles during the period from January 26 to 30, 2016 at the 17 pressure levels between 1000 and 10 hPa with a latitude step of 2.5°. For the calculations, the following formula was used:

$$F_3 = \frac{1}{R_3} \frac{pv}{T} r_3$$

where $F_3$ is the ozone flux (µg m$^{-2}$ s$^{-1}$), $r_3$ is the ozone VMR (ppbv), $v$ is the meridional wind speed (m s$^{-1}$), positive values correspond to the south wind and the ozone flux from the south to the north, $p$ is the air pressure (hPa), $T$ is the air temperature (K) and $R_3$ = 173 J kg$^{-1}$ K$^{-1}$ is the gas constant for ozone. At each pressure level, the ozone inflow in the region centered at the latitude $\varphi_0$ was calculated as the difference between the ozone fluxes at the latitudes $\varphi_0 - 2.5°$ and $\varphi_0 + 2.5°$.

It can be seen from Fig. 7a, that in the longitude segment of the OMH manifestation, the north wind dominated in the stratosphere, while the south wind dominated in the troposphere that is in tune with the results of trajectory analysis (Fig.6a, 6b). The average speed of the north wind in the stratosphere occurred between 60° N and 75° N during the period from 26 to 30 January 2016 was -12.8 m/s, while that of the south wind in the troposphere amounted to +7.4 m/s. Taking into account the opposite latitudinal gradients of ozone in the stratosphere and in the troposphere, the ozone transfer in the indicated altitude regions contributed to a decrease in total ozone content over the blocked region.

Fig. 7b presents the results of calculations of the inflow and outflow of ozone at different levels over the Northern extratropics in the end of January 2016. It can be seen that in the stratosphere the ozone outflow took place over all latitudes north of 55° N, while near 60° N it embraced the entire range of altitudes. The local extreme in the ozone VMR decrease (-275 μg/m$^3$s) was reached in the lower stratosphere between 65° N and 70° N and coincided with the latitudinal OMH position. An analysis of Fig. 7a shows that the extreme was caused by dynamical reason, because the ozone outflow from this region essentially exceeded the ozone inflow in it.

The evolution of the vertical distribution of the ozone anomalies within the blocking anticyclone in the second half of January and in the first ten days of February 2016, is shown in Fig. 8a. It can be seen that the negative ozone anomalies over the blocked region prevailed from the surface layer to the upper stratosphere. The greatest ozone decrease occurred in the lower stratosphere, however, near the local minima, the strong ozone decrease spread up till the altitudes of the upper stratosphere. The local minima of the ozone anomalies were manifested near the level of 70 hPa on January 21, 26 and 29, 2016, reaching the values of -1456, -1403 and -1454 ppbv (-49, -47, -47% relative the long-term values of the ozone VMR for this level and corresponding date during the period 2003-2015). In the lower stratosphere over the blocked region, negative ozone anomalies were accompanied by negative temperature anomalies (cf. Fig. 8a with Fig. 8b). In particular, the extreme negative ozone anomalies displayed at the end of January 2016, were accompanied by the temperature anomalies below -30 °C. As noticed earlier, such a relationship between the temperature and ozone variations is characteristic for ascending motions of air in the lower stratosphere.

However, a decrease in stratospheric temperature and accompanying decrease in ozone VMR can be in part assigned to the transport of extremely cold- and ozone-depleted air from the inner Arctic into the blocked region (Fig.6a) due to the displacement of the polar vortex.

In the first ten days of February 2016, an increase in stratospheric temperature (Fig. 8b) and a concomitant increase in stratospheric ozone (Fig.8a) were caused by minor sudden stratospheric warming.

*4.6. Chemical mechanism of ozone mini-hole over Siberia in winter 2015/2016*

The chemical mechanism of the formation of OMHs is associated with the destruction of ozone in catalytic reactions with halogens in the presence of PSCs, regularly observed in winter and spring in the Arctic lower stratosphere. The PSCs are acid droplets formed at temperatures below -78°C (PSC type I) and ice particles formed at temperatures below -85°C (PSC type II). Under extremely low stratospheric temperatures, occurring during polar night conditions, heterogeneous chemical reactions on the surface of PSCs promote the conversion of inert halogen-containing species into active forms of chlorine and bromine. The following illumination of atmosphere by solar radiation triggers a chain of photochemical reactions of ozone destruction with the participation of halogens as catalysts (Solomon 1999).

As noted earlier, in the second half of January and the first ten days of February 2016 in the lower stratosphere above the blocked region, temperatures below -78 °C were detected, fallen below -85 °C during the period from January 25 to 30 (see Fig. 5c and Fig. 8b). Low temperatures covered a fairly wide region of altitudes. Thus, on January 27-28, 2016, the temperatures below -78 °C were detected between the pressure levels from 100 up to 10 hPa (16 - 31 km), while those below -85 °C were found between 80 and 30 hPa (18 - 24 km). The minimum temperature above the blocked region was detected on January 28, 2016 at the pressure level of 30 hPa, where temperature has dropped down to -89.8 °C. Such conditions favored for the PSC type I and type II formation. It is worth noting that at the latitude of the central part of the OMH, observed over Western Siberia in early 2016 (~ 67° N), the polar night ends in mid-January.

Although the main reason for PSCs in the Arctic to form is a strong decrease in stratospheric temperature inside the polar vortex under polar night conditions, there are a number of additional factors affecting a thermal regime of the stratosphere, in particular, the displacements of the polar vortex. The presence of planetary waves, synoptic-scale waves and mountain waves, affecting the tropopause height can also contribute to decrease in the lower stratospheric temperature that favors PSCs to form (Teitelbaum et al., 2001; Fueglistaler et al., 2003; Kohma and Sato, 2013). The results obtained in this study show that increase in the tropopause height over the blocked region is indeed accompanied by a decrease in the lower stratospheric temperature (cf. Fig 5a with Fig 5c). Fromm et al. (2003) concluded that tropospheric, synoptic-scale flow perturbations are the primary forcing mechanism for Arctic PSC formation. In addition, the degree of pollution of the high-latitude stratosphere affects the processes of heterogeneous nucleation when PSCs form and grow.

Unfortunately, during the period from January 28 till March 13, 2016 the CALIOP measurements were absent. In addition, despite the sufficient number of measurement days in January, not every subsatellite track crossed the ozone mini-hole. Fig. 9a shows the total ozone field on January 27, 2016 over Siberia with the superimposed ground-based projection of the CALIPSO orbit, passed through OMH that day. It can be seen from Fig. 5a that January 27, 2016 is one of the days when TCO inside OMH reached its lowest values.

The altitude distribution of aerosol types detected by the CALIOP instrument along the subsatellite track (Fig. 9b) indicates that despite a variety of aerosol subtypes observed in the tropospheric part of OMH, the PSC aerosol dominated in its stratospheric part. The CALIOP algorithm categorizes PSCs into four groups: super-cooled ternary solution (STS) droplets, mixture of nitric acid trihydrate (NAT) particles and STS droplets, with a low content of NAT particles (hereafter NAT), mixture of NAT and STS, with a high content of NAT (enhanced NAT or NATe) as well as ice crystals (Pitts et al., 2009). Fig. 9c shows the altitude distribution of the composition of PSCs, detected by the CALIOP instrument along the subsatellite track on 27 January 2016. An important feature of the PSC distribution along the orbit projection is the fact that the bulk of the PSCs is concentrated in the OMH area (compare Fig. 9a with Fig. 9c), while outside OMH PSCs are practically absent. The latter feature testifies in favor of the existence of the chemical destruction of ozone in heterogeneous reactions with halogen-containing compounds on the surface of PSCs, occurred on January 27, 2016.

On the whole, in the composition of PSCs, detected by CALIOP in the stratosphere within the Siberian OMH on January 27, 2016, in approximately equal proportions STS, NAT and NATe are contained, with an insignificant percentage of ice crystals (Fig. 10). This result is in tune with the long-term annual-mean PSC composition characteristic for the Arctic region (Pitts et al. 2011). Along with PSC aerosol, in the upper troposphere and in the vicinity of the tropopause layer ice crystals simultaneously existed (Fig. 9c), that could be related to the formation of the cirrus clouds due to blocking high (Kohma and Sato, 2013). It is generally considered that the most common type of the PSC aerosol during the Arctic winter is STS and NAT (e.g. Wegner et al 2012). The obtained results agree with this conclusion. It follows from Fig. 10 that in total, the contribution of STS and NAT over the blocked region reaches nearly 70%.

It should be noted that chlorine activation occurs more massively on liquid STS droplets, not on solid particles (NAT or ice) (Tritcher et al., 2021). The activation begins at air temperatures below -78 °C when liquid particles start to grow, taking up ambient $HNO_3$ and $H_2O$ (Carslaw et al., 1994; Peter, 1997). Simultaneously with chlorine activation, chlorine deactivation occurs which depend on the processes of denitrification. Sedimentation of large NAT particles promotes removing $HNO_3$ from the lower stratosphere, that suppresses chlorine deactivation in the reactions with reactive nitrogen and thus prolongs residing of active chlorine in the atmosphere. It can be expected that the time response of ozone on chemical factors is longer than that on dynamical ones.

The top panel in Fig. 11 shows the PSCs, detected by CALIOP at different altitudes, over the control points on the subsatellite track in Fig. 9a and the ozone profiles corresponding to them. In the same figure the ozone profile, calculated by averaging the ozone vertical distributions at the beginning (79.02° N, 118.99° E) and at the end (51.31° N, 79.46° E) of the subsatellite track is also shown. Such an average profile is used as a reference one, since ozone distribution strongly depends on latitude. It can be seen from Fig. 11 that for the profiles inside the ozone mini-hole, the ozone VMR is, on the whole, noticeably lower than outside it. The number of PSCs as well as the thickness and effective height of the PSC layer essentially change along the trajectory. From the beginning to the end of the subsatellite track inside the ozone mini-hole, an increase of PSC altitudes in the lower stratosphere is noted, accompanied by a decrease in the aerosol content near the tropopause layer. Also, from the edge of the ozone mini-hole toward its center, a widening of the PSO layer is characteristic. The changes in the PSC profiles are accompanied by rising the height positions of ozone VMR extrema in the stratosphere. However, because the differences between current and reference ozone profiles are most noticeable in the upper stratosphere (up to 40% at the 5 hPa level), it is unlikely that they are associated with the destruction of ozone due to reactions on PSCs.

In the lower panel of Fig. 11 a stratospheric region of PSC manifestation is shown in an enlarged form. In addition to current and reference ozone profiles each plot also presents the current vertical distributions of PSC composition over the control points. It can be seen, that PSC composition depends on altitude and also varies along the subsatellite track. Among the most notable changes is an increase in the effective height of STS along the trajectory, and also an increase the percentage of NATe in the inner regions of OMH. According to the CALIOP data, the largest number of PSCs on January 27, 2016 are found over the control point (68.83°N, 92.16°E) between the pressure levels 100 and 50 hPa (~17-21 km).

In the inner region of the ozone mini-hole, ozone mixing ratios between the pressure levels 200 and 50 hPa (~13-21 km) are systematically lower compared to those for the reference ozone profile. The maximum difference of ozone mixing ratios (386 ppbv) was detected over the control point (68.83° N, 92.16° E) in the vicinity of the pressure level of 100 hPa. It is characteristic that the lowest total ozone content along the entire trajectory (of about 240 DU) on January 27, 2016 was detected close this reference point. This circumstance can be explained by the fact that a chemical decrease in stratospheric ozone is also manifested in changes in TCO that in turn can be considered as proof of the manifestation of the action of the chemical mechanism of the formation of the OMH over western Siberia in 2015/2016.

Within the pressure levels 200-20 hPa, the differences between ozone abundance calculated for the ozone profiles over the control points inside the ozone mini-hole and that for the reference ozone profile are found to be in the range -12 to -25 DU. Taking into account the fact that total ozone content, calculated using reference ozone profile is 307 DU, the estimates of the chemical destruction of ozone in the ozone mini-hole lie in the range from -4 to -8%.

**5. Conclusion**

An analysis of the characteristics and mechanisms of the formation of the deep OMH over Western Siberia during winter 2015/2016 was carried out. The depletion in TCO, reaching -140 DU in the end of January, was caused mainly by a strong decrease in ozone mixing ratio in the lower stratosphere, reaching 50% near the pressure level of 70 hPa. The formation of the OMH was mainly explained by the dynamic factors including an increase in the tropopause height and advection in the troposphere of ozone-depleted subtropical air, which were conditioned by atmospheric blocking. Also, an important factor of the OMH formation was an ozone decrease due to increased outflow of ozone from the area of the OMH in the stratosphere. Extremely low stratospheric temperatures over the Northern high latitudes and also dynamic perturbations in the troposphere, resulting in elevation of the tropopause and additional decrease in stratospheric temperature, favored the formation of PSCs within OMH region. However, the contribution of chemical destruction of stratospheric ozone, due to catalytic reactions with halogen-containing compounds on the surface of PSCs to the OMH formation, did not exceed 10% at that.

**Acknowledgements** The authors gratefully acknowledge the AIRS, CALIOP, NCEP/NCAR reanalysis and HYSPLIT teams for their effort in making the data available. This work was supported by the Russian Science Foundation (project 19-17-00240).

**Figure captions**

**Figure 1.** Day-to-day variations of: (a) the blocking index in the longitudinal segment of (55°-140° E), (b) total column ozone over the region of (55°–85° N, 55°-140 °E) during winter 2015/2016. Each symbol in (a) marks a day with atmospheric blocking and a longitude at which corresponding blocking reveals itself.

**Figure 2.** Spatial distributions for January 26-30, 2016 of: (a) 500-hPa geopotential height ($H$500), (b) the anomalies of total column ozone ($\Delta O_3$).

**Figure 3.** Latitude-pressure cross-sections over the longitude segment of 60°-110° E of: (a) ozone VMR anomalies for January 26-30, 2016, (b) the long-term averages of ozone VMR for January 26-30, 2003-2015.

**Figure 4.** Spatial distributions for January 26-30, 2016 of: (a) the wind vectors at the 500-hPa pressure level, (b) the tropopause height anomalies (color fill) and $H$500 (pale isolines, see also Fig. 2a).

**Figure 5.** Day-to-day variations in January and February 2016 of: (a) the total column ozone ($O_3$) and the tropopause height ($H_{trop}$) over the blocked region (60°-75° N, 70°-100° E), (b) the meridional wind over the western periphery of the block (50°-75° N, 60°-75° E, $v_{S \to N}$) and that over the eastern periphery of the block (50°-75° N, 70°-100° E, $v_{N \to S}$) and (c) the air temperature over the blocked region and the meridional wind over the north of the blocked region (70°-80 N, 70°-100° E) at the 50-hPa pressure level. Red (blue) lines in (c) denotes temperature thresholds of PSC type I (type II) formation, respectively.

**Figure 6.** The sets of 54 backward trajectories, arriving in the OMH region, bounded by 65°-70° N and 85°-93° E, on January 27, 2016 at the heights of 20 km (a) and 5 km (b). Matrix type of trajectories calculation was utilized.

**Figure 7.** Latitude-pressure cross-sections over the longitude segment 60°-75° E of: (a) the meridional wind, (b) the regions of ozone inflow (warm colors) and ozone outflow (cold colors). Averaging for January 26-30, 2016.

**Figure 8.** Time-pressure cross-sections over the blocked region (60°-75° N, 70°-100° E) in January and February 2016 of: (a) the ozone VMR anomalies, (b) the air temperature anomalies. In (b) the atmospheric region with the air temperatures below –78 °C is shown by light dashed line, while that with the air temperatures below –85 °C - by purple solid line.

**Figure 9. (a)** The total column ozone field over Siberia on January 27, 2016 according to AIRS data (toned), white line shows ground projection of the CALIPSO orbit on January 27, 2016 between 20:59:36 and 21:08:02 UTC, white circles are the control points coinciding with the coordinate stamps on the X-axes in (b) and (c). (b) - The vertical distribution of aerosol subtypes. (c) - the PSC composition along the subsatellite track shown in (a); thick dotted line denotes the tropopause height.

**Figure 10.** Percentage of the PSC aerosol composition in the stratosphere as a whole, obtained by averaging the CALIOP measurements over the entire subsatellite track within ozone mini-hole (see also Fig. 9a).

**Figure 11.** PSCs, detected by CALIOP at different altitudes on January 27, 2016 over the control points inside the ozone mini-hole: (a) 74.28°N, 101.50°E, (b) 68.83°N, 92.16° E, (c) 64.09°N, 86.41°E and (d) 57.22°N, 82.44°E (tone, upper X-axes) and the ozone VMR profiles, obtained by AIRS in the vicinity of each control point (blue curves, bottom X-axes), red curves depict the reference ozone profile, calculated by averaging the profiles over the control points on the edges of ozone mini-hole: 79.02°N, 118.99°E and 51.31°N, 79.46°E, thick dotted lines denote tropopause position over each control point. (e-h) The same as (a-d) but enlarged for the lower stratosphere, the PSC composition categorized by colors, the reference ozone profile is shown by dotted lines.

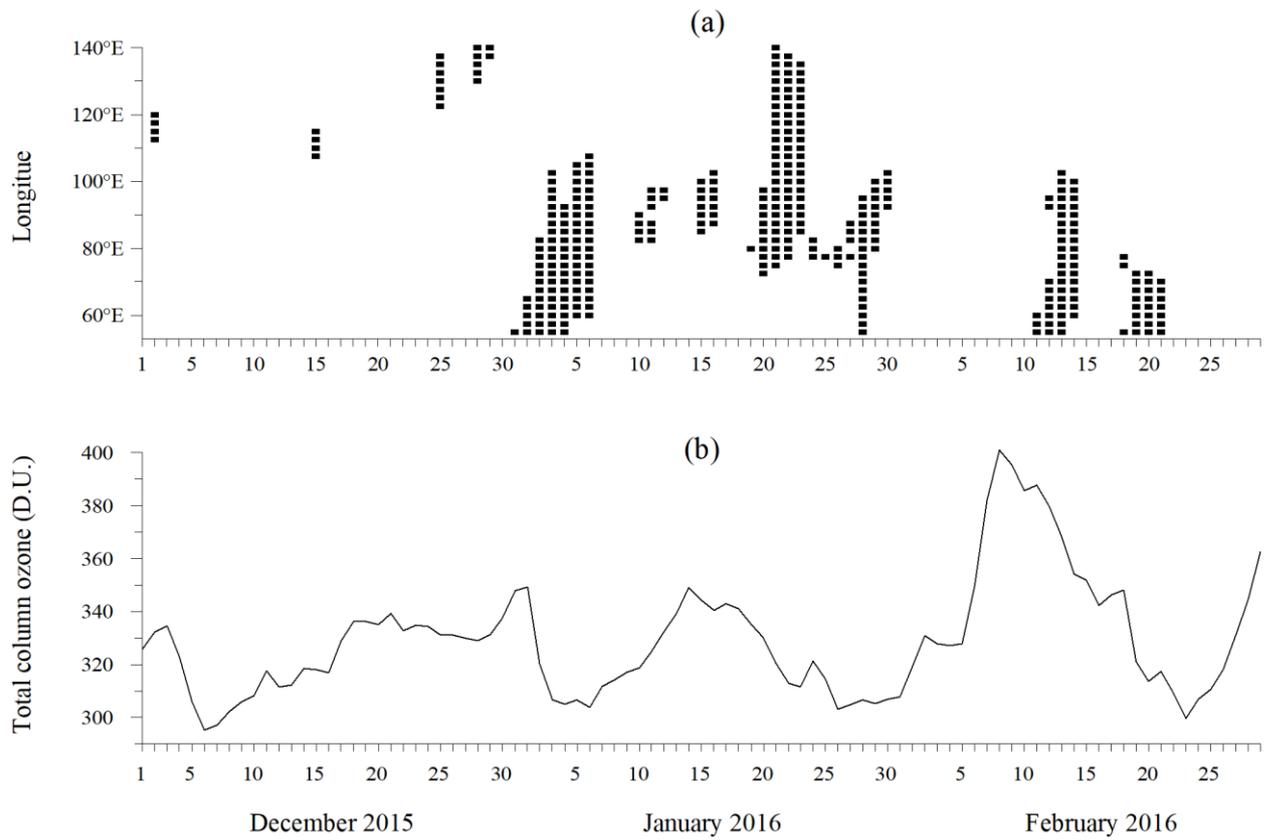

**Figure 1.** Day-to-day variations of: (a) the blocking index in the longitude segment (55°-140° E), (b) total column ozone over the region of (55°–85° N, 55°-140 °E) in winter 2015/2016. Each symbol in (a) marks a day with atmospheric blocking and a longitude at which corresponding blocking reveals itself.

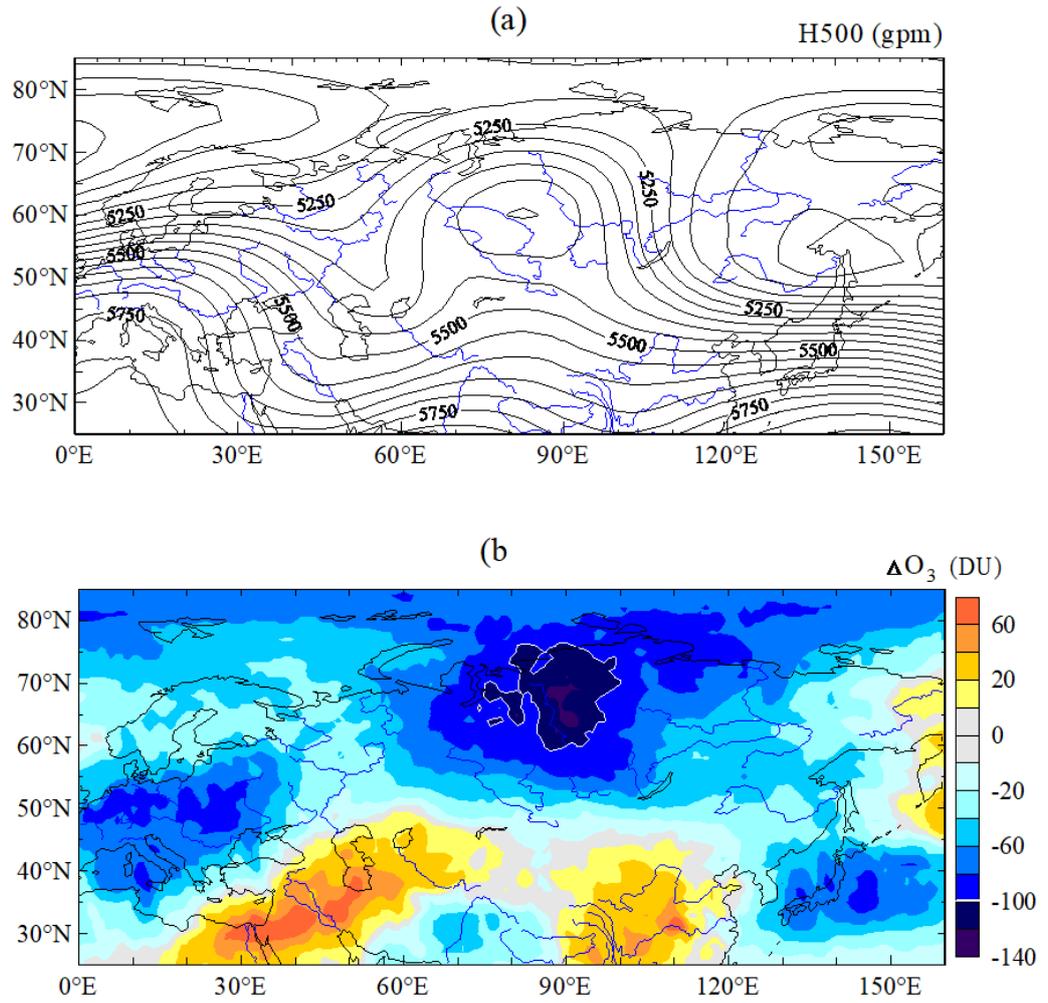

**Figure 2.** Spatial distributions for January 26-30, 2016: (a) 500-hPa geopotential height ($H$500), (b) the anomalies in total column ozone ($\Delta O_3$).

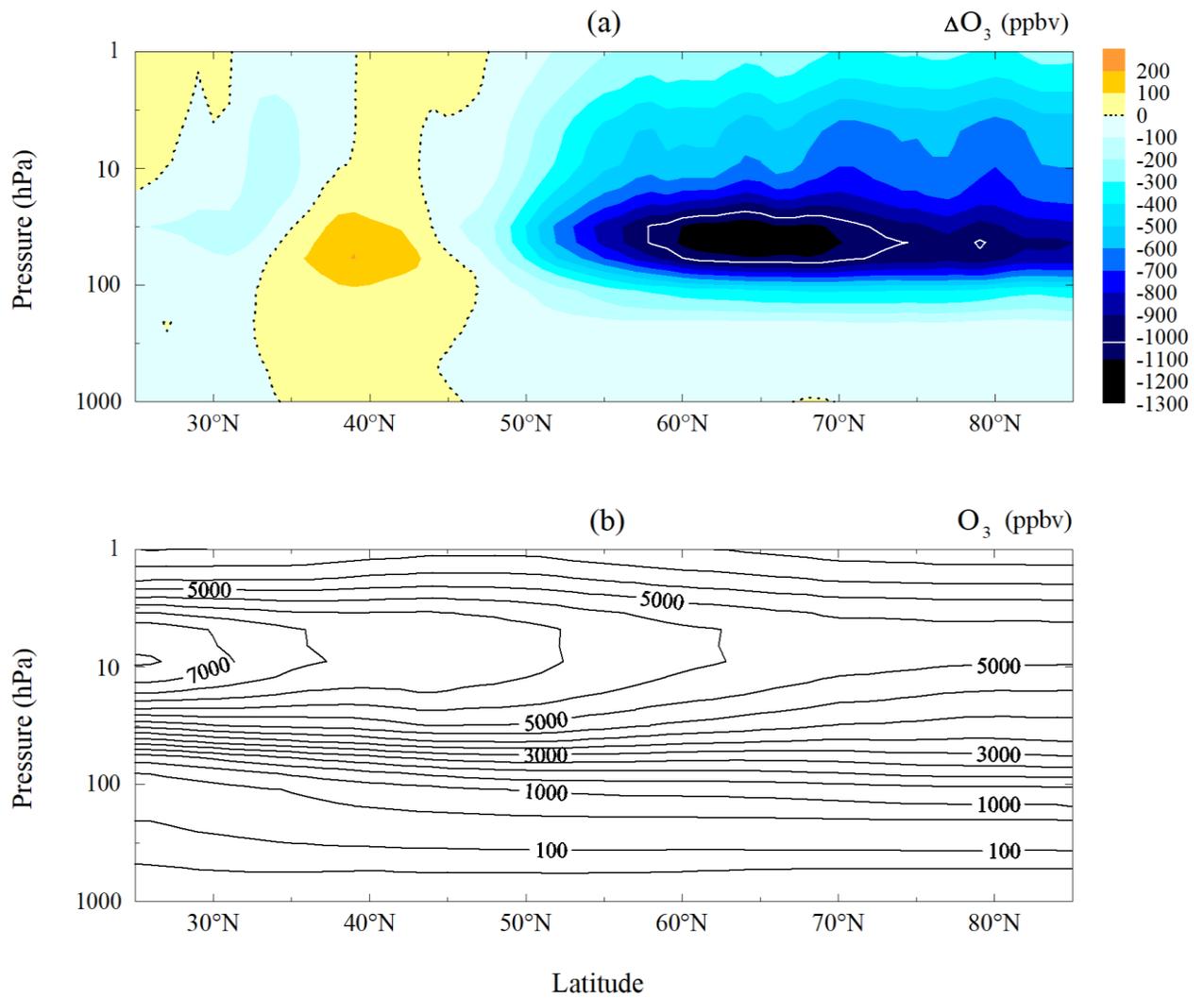

**Figure 3.** Latitude-pressure cross-sections over the longitude segment of 60°-110° E: (a) ozone VMR anomalies for January 26-30, 2016, (b) the long-term averages of ozone VMR for January 26-30, 2003-2015.

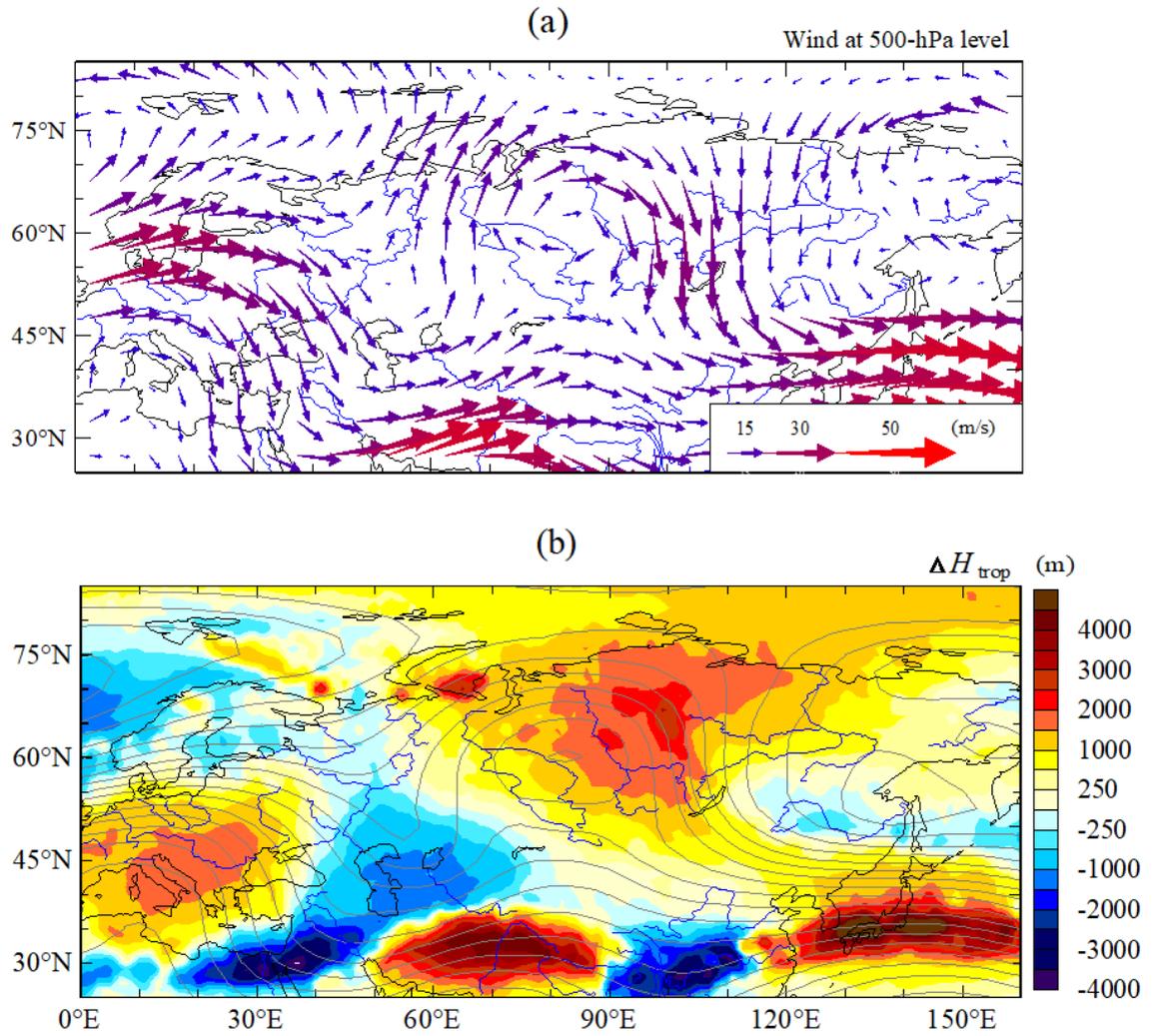

**Figure 4.** Spatial distributions for January 26-30, 2016: (a) the wind vectors at the 500-hPa pressure level, (b) the tropopause height anomalies (color fill) and $H500$ (pale isolines, see also Fig. 2a).

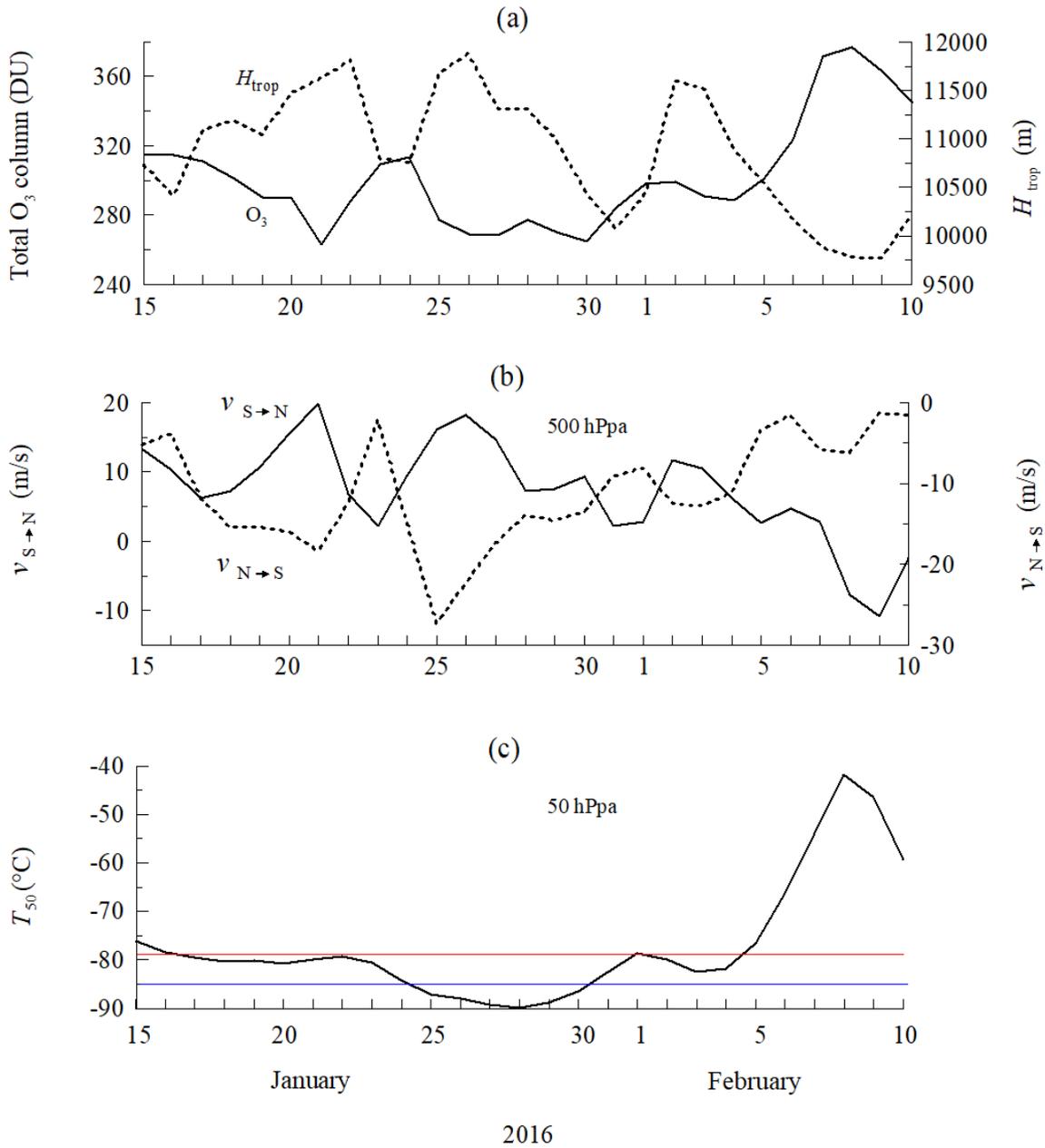

**Figure 5**. Day-to-day variations in January and February 2016: (a) the total column ozone ($O_3$) and the tropopause height ($H_{trop}$) over the blocked region (60°-75° N, 70°-100° E), (b) the meridional wind over the western periphery of the block (50°-75° N, 60°-75° E, $v_{S \to N}$) and that over the eastern periphery of the block (50°-75° N, 70°-100° E, $v_{N \to S}$) and (c) the air temperature over the blocked region at the 50-hPa pressure level. Red (blue) lines in (c) denotes temperature threshold of the PSC type I (type II) formation, respectively.

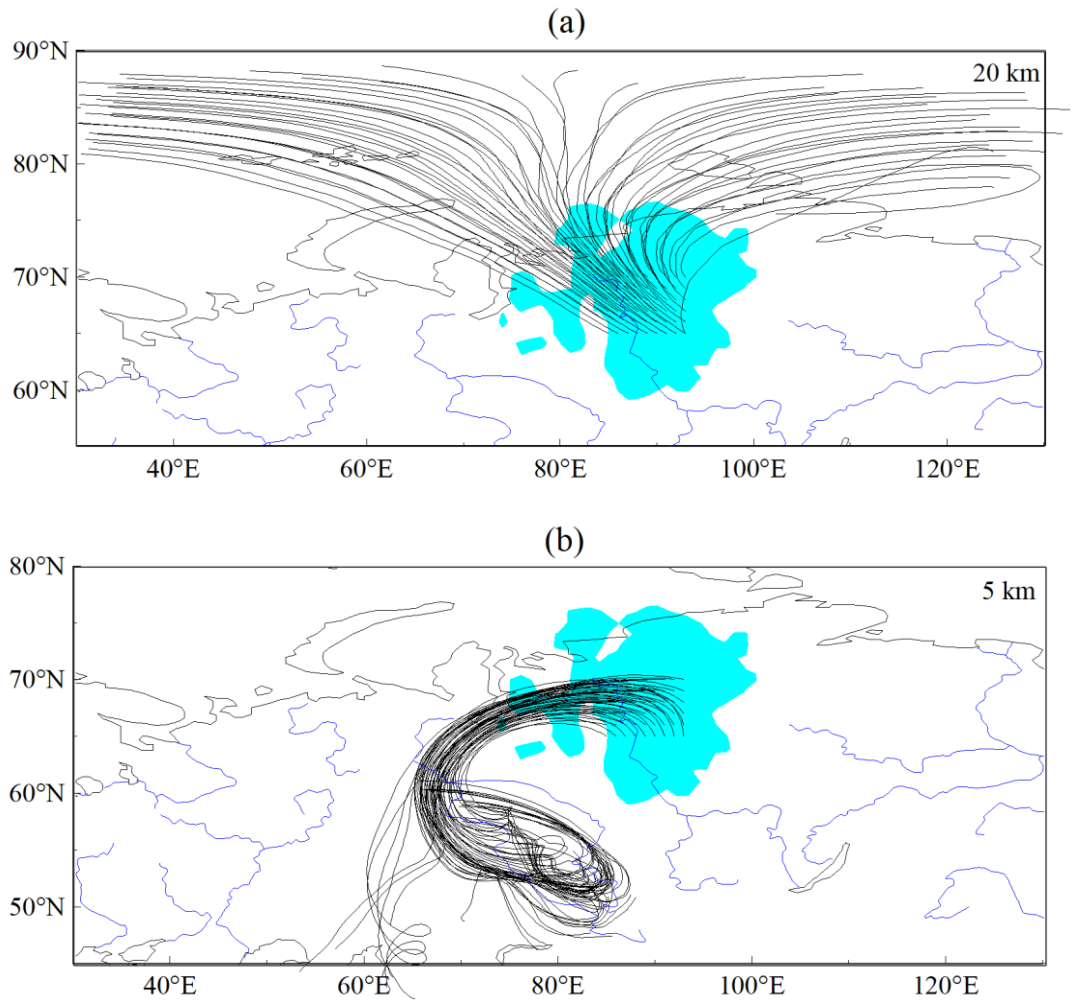

**Figure 6.** The ensembles of 54 backward trajectories, arriving in the region bounded by 65°-70° N and 85°-93° E on January 27, 2016 at the heights of 20 km (a) and 5 km (b). Matrix type of trajectories calculation was utilized.

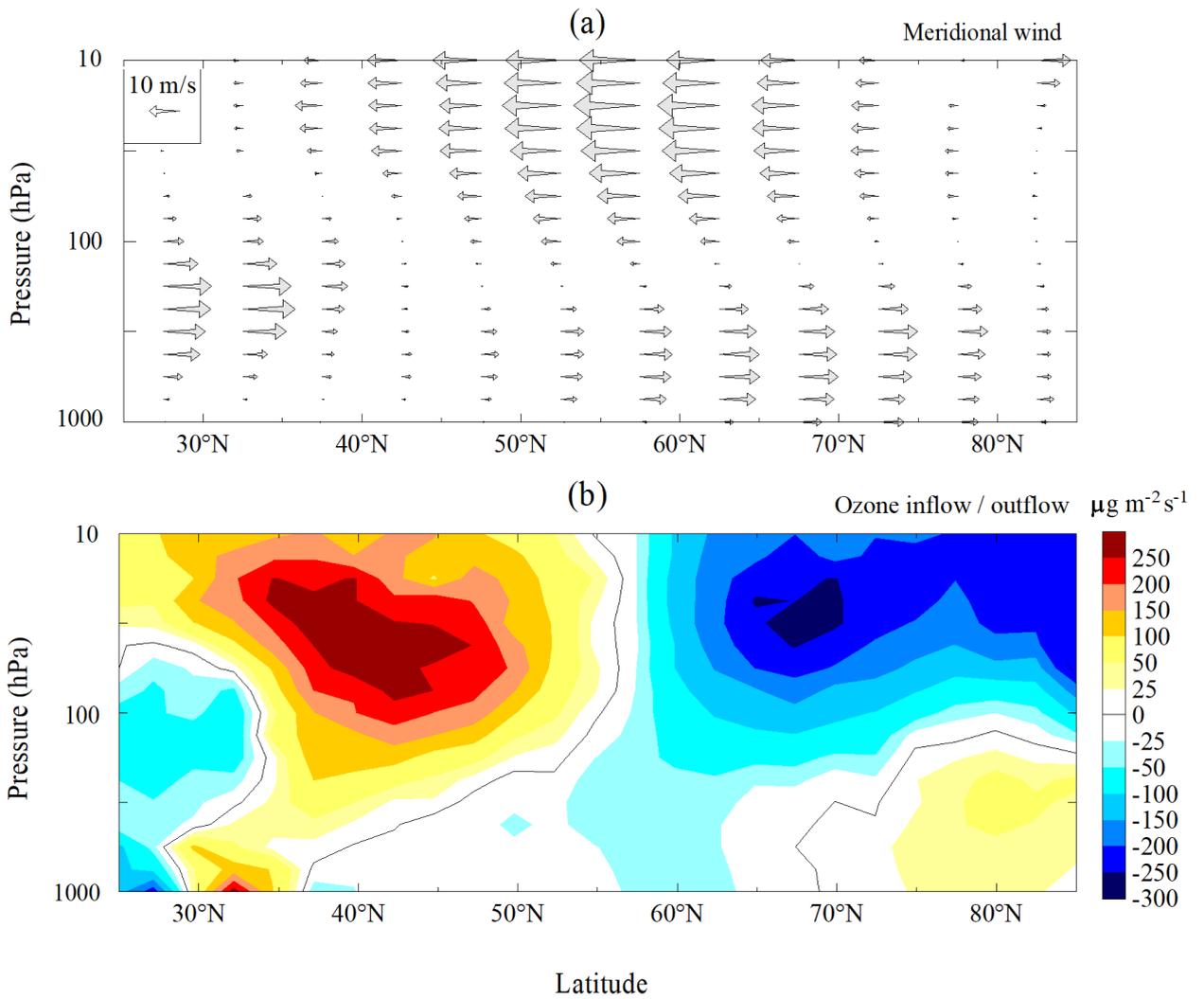

**Figure 7**. Latitude-pressure cross-sections over the longitude segment 60°-75° E for January 26-30, 2016: (a) meridional wind, (b) the regions of ozone inflow (warm colors) and ozone outflow (cold colors).

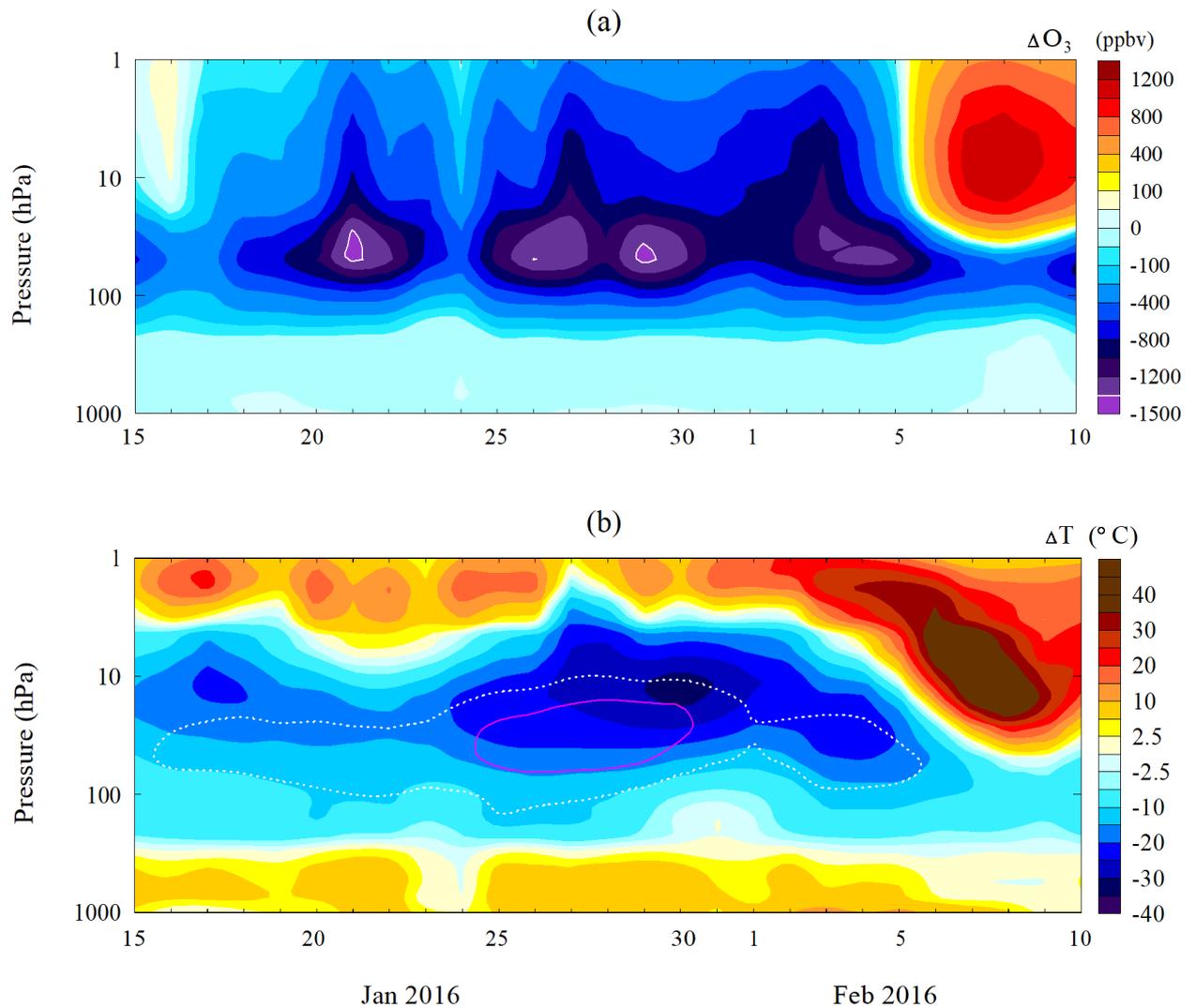

**Figure 8.** Time-pressure cross-sections over the blocked region (60°-75° N, 70°-100° E): (a) the ozone VMR anomalies, (b) the air temperature anomalies between January 15 and February 10, 2016. In (b) the atmospheric region with the air temperatures below –78 °C is shown by light dashed line, while that with the air temperatures below –85 °C - by purple solid line.

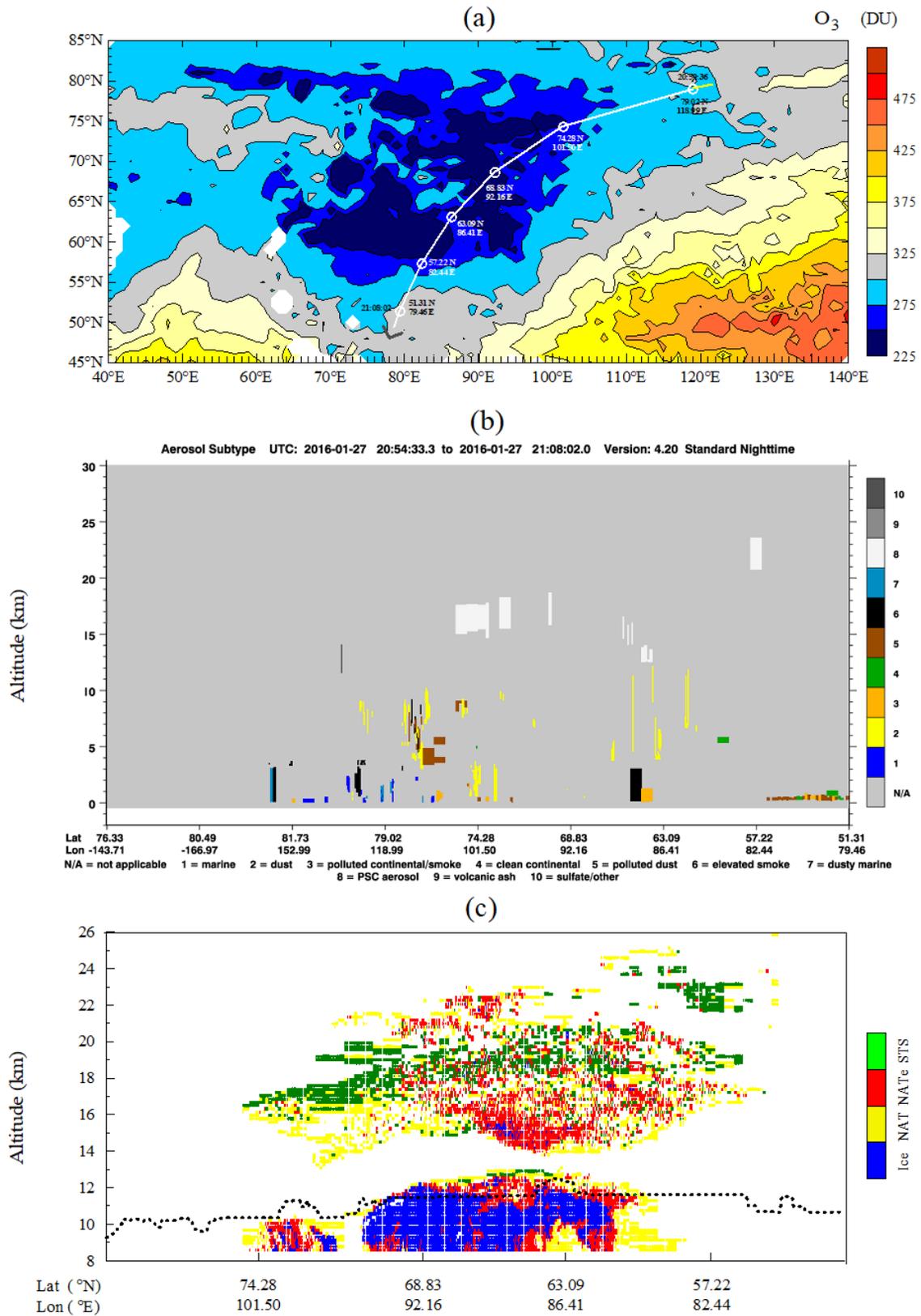

**Figure 9**. (a) The total column ozone field over Siberia on January 27, 2016 according to AIRS data (toned), white line shows ground projection of the CALIPSO orbit on January 27, 2016 between 20:59:36 and 21:08:02 UTC, white circles are the coordinate stamps, shown also on the X-axes in (b and c). (b) The vertical distribution of aerosol subtypes. (c) The PSC composition along the subsatellite track shown in (a). The thick dotted line in (c) denotes the tropopause height.

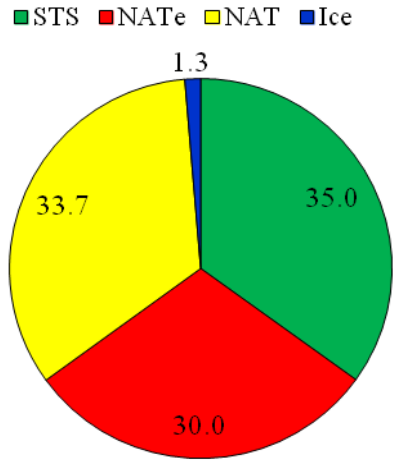

**Figure 10.** Percentage of the PSC aerosol composition in the stratosphere as a whole, obtained by averaging the CALIOP measurements over the entire subsatellite track within ozone mini-hole (see also Fig. 9a).

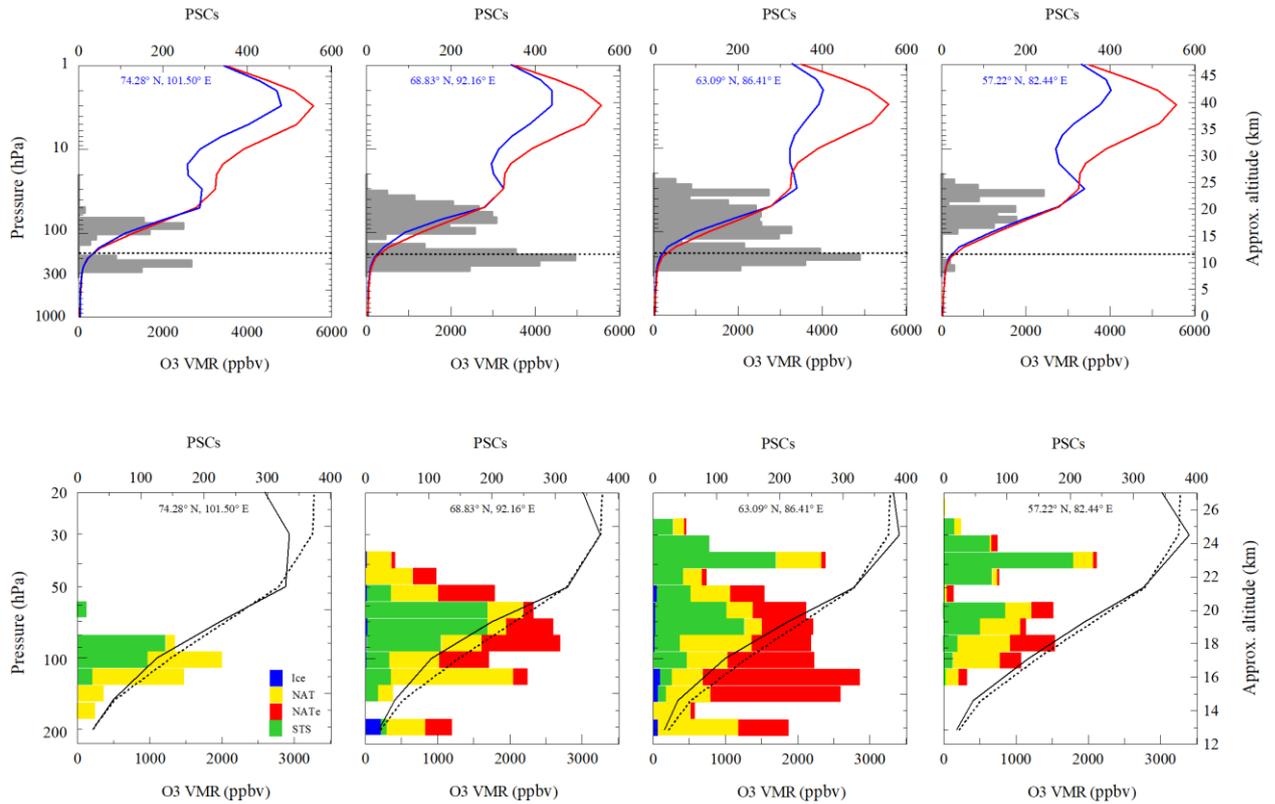

**Figure 11.** PSCs, detected by CALIOP at different altitudes on January 27, 2016 over the control points inside the ozone mini-hole: (a) 74.28°N, 101.50°E, (b) 68.83°N, 92.16° E, (c) 64.09°N, 86.41°E and (d) 57.22°N, 82.44°E (tone, upper X-axes) and the ozone VMR profiles, obtained by AIRS in the vicinity of each control point (blue curves, bottom X-axes), red curves depict the reference ozone profile, calculated by averaging the profiles over the control points on the edges of ozone mini-hole: 79.02°N, 118.99°E and 51.31°N, 79.46°E, thick dotted lines denote tropopause position over each control point. (e-h) The same as (a-d) but enlarged for the lower stratosphere, the PSC composition categorized by colors, the reference ozone profile is shown by dotted lines.